# Flow-PIN: A Two-Stage Power-Flow-Guided Method for System-Wide Multivariate Profile Inpainting in Distribution Networks

Zhenghao Zhou, *Student Member*, *IEEE*, Yiyan Li, *Member*, *IEEE*, Yike Guo, Xinyi Ma, Zheng Yan, *Senior Member*, *IEEE*, and Mo-Yuen Chow, *Life Fellow, IEEE*

***Abstract*—High-quality system measurement data is critical for power distribution system operation. As deep generative models (e.g., GAN, Diffusion, etc.) have been widely studied to solve the missing data restoration problem to enhance the data quality, their results may look "realistic" but not sufficiently "accurate" due to lacking physical guarantees. To address this limitation, a two-stage physics-guided framework, Flow-PIN, is proposed in this paper for system-wide multivariate profile inpainting. The first stage employs a conditional flow matching model, conditioned on topological and correlation graphs, to generate candidate values. A physical penalty is integrated into the loss function to constrain the generative vector field based on grid physical laws. The second stage introduces a topology-aware power-flow-guided refiner that utilizes Laplacian positional encoding to inject topology information into node embeddings. By coupling alternating current power flow equations with a differentiable correlation alignment mechanism, this refiner further corrects numerical deviations. Evaluations on an active distribution network dataset benchmark the proposed framework against ten representative baselines. The results show that Flow-PIN achieves high-fidelity profile inpainting across three dimensions: maximizing numerical accuracy, capturing temporal fluctuation, and preserving spatial topological correlations.**

## I. Introduction

MODERN distribution networks heavily rely on high-quality system measurement data to achieve accurate situational awareness tasks, such as state estimation, load forecasting and fault identification, so that the operation challenges brought by stochastic Distributed Energy Resources (DER) can be alleviated. However, temporary equipment failures or communication disruptions lead to missing data, which severely compromises data quality. Therefore, recovering missing data from incomplete observations to construct high fidelity datasets is of paramount importance to power distribution system operations.

Existing missing data restoration methods can be broadly categorized into statistical-based approaches and machine learning methods. Statistical-based approaches infer missing values by modeling the inherent statistical patterns of power data. Classic methods, including autoregressive models and linear regression [1], rely on local temporal similarity but deteriorate rapidly during continuous data loss [2]. To capture global redundancy, advanced statistical frameworks frequently formulate the recovery task as low-rank matrix or tensor completion. By mapping high-dimensional measurements into lower-dimensional statistical subspaces, these methods effectively reconstruct missing data utilizing techniques such as Bayesian estimation [3], physical constraint regularization [4], and advanced tensor decompositions [5], [6]. Despite their effectiveness and theoretical tractability, both classic and low-rank statistical models usually depend on predefined smoothness, low-rank or physically motivated structural assumptions. Although these assumptions can be strengthened through Bayesian estimation, regularization, and domain-knowledge constraints, they may still limit model flexibility when reconstructing highly complex and nonlinear spatiotemporal patterns in networks.

With the rapid advancement of artificial intelligence, machine learning methods, particularly deep learning models, have gained significant attention. Razavi-Far et al. propose a hybrid data imputation method based on correlation connected clustering and genetic algorithm optimization [7]. Wu et al. extract low dimensional features from spatiotemporal graphs via network embedding and random walks, utilizing regression neural networks to resolve diverse missing data patterns [8]. Xu et al. develop a recovery method for electrical load curves driven by bidirectional gated recurrent units [9]. Their method utilizes dynamic time warping techniques to match and graft similar curve segments from a historical database to restore periodic trends. Li et al. proposed a federated learning approach, which integrates data from multiple wind farms to achieve precise recovery and introduces a zero-trust federated learning framework to ensure privacy [10]. Building upon these predictive machine learning techniques for spatiotemporal feature extraction, researchers have further investigated methods to model complex data distributions.

Within the field of deep learning models, advanced generative models demonstrate remarkable capabilities in capturing complex data distribution, such as data generation [11] and super resolution [12]. Generative Adversarial

This work was supported by National Natural Science Foundation of China under Grant 52307121. (Corresponding author: Yiyan Li.)

Zhenghao Zhou, Yiyan Li, Yike Guo, Xinyi Ma are with the College of Smart Energy, Shanghai Non-Carbon Energy Conversion and Utilization Institute, and Key Laboratory of Control of Power Transmission and Conversion, Ministry of Education, Shanghai Jiao Tong University, Shanghai, 200240, China. (e-mail: zhenghao.zhou@sjtu.edu.cn, yiyan.li@sjtu.edu.cn, yike.guo@sjtu.edu.cn, xinyi.ma@sjtu.edu.cn).

Zheng Yan is with the Key Laboratory of Control of Power Transmission and Conversion, Ministry of Education, and the Shanghai Non-Carbon Energy Conversion and Utilization Institute, Shanghai Jiao Tong University, Shanghai 200240, China. (e-mail: yanz@sjtu.edu.cn)

Mo-Yuen Chow is with the Global College, Shanghai Jiao Tong University, Shanghai, 200240, China. (email: moyuen.chow@sjtu.edu.cn)

Networks (GAN) find extensive application in restoring missing data [13]. For instance, Fang et al. and Li et al. respectively propose spatiotemporal feature extraction networks and coarse to fine adversarial architectures [14], [15]. These methods enable the precise recovery of variable length data segments without the prerequisite of complete training datasets or fixed input formats. Sun et al. integrate an improved state transition algorithm with GAN to optimize interval allocation and generate realistic recovered data [16]. Advancing probabilistic frameworks, Dong et al. introduce a diffusion model equipped with hierarchical consistency regularization to specifically address stratified imputation challenges in complex time series [17]. Furthermore, transformer architectures and large language models establish novel paradigms for data restoration. Hu et al. capture long range dependencies by reformulating missing data recovery as a classification task via bidirectional encoder representations [18]. Meanwhile, smart meter data can be encoded into text sequences. By harnessing the prior knowledge of pretrained language models through prompt engineering, this strategy achieves accurate data recovery utilizing minimal samples [19].

Despite the success of deep generative models in time-series imputation, it should be noted that these models focus on learning the inherent data distribution of the training samples to create new ones. As such, the missing data restoration results from these models may look “realistic” but not “accurate”. On the other hand, supervised learning models and physics-based methods focus on point-wise precision. However, they may lead to conservative and over-smoothed restoration results without capturing the high-frequency details. In other words, the results from supervised learning models and physics-based methods may be numerically “accurate” but not “realistic”.

In this paper, we propose a Flow-based Profile Inpainting Network (Flow-PIN), combining the advantages of both generative models in learning data distribution and regression models in improving point-wise accuracy. Flow-PIN is a two-stage physics-guided framework designed to inpaint missing segments for system-wide multivariate power system profiles. The first stage utilizes a physics-guided Conditional Flow Matching (CFM) model, which is a generative mapping method derived from optimal transport theory, to establish a statistical foundation, mapping random noise to empirical distribution. The second stage introduces a Topology-aware Power-flow-guided Refiner (TPR) to further improve numerical precision. By embedding physical laws directly into the learning objectives of both stages, the framework achieves high-fidelity profile inpainting across point-wise, temporal, and spatial dimensions.

The contributions are summarized as follows:

1. A two-stage power-flow-guided profile inpainting framework, named Flow-PIN, is proposed for system-wide multivariate power-system profiles. The framework decouples distribution-aware candidate generation from physics-aware refinement: the first stage learns realistic spatiotemporal profile distributions, while the second stage corrects local numerical deviations under grid-physical constraints. This design improves point-wise accuracy and physical consistency while preserving temporal fluctuations and spatial topological correlations.

2. A conditional flow matching model is developed to generate candidate values for multivariate blind zones. By utilizing conditional information derived from topological and correlation graphs, this generative model captures the spatiotemporal dynamics of the distribution network. A dynamically weighted physical loss is further incorporated to guide the generative path toward a physically feasible domain without suppressing distribution learning.

3. A topology-aware power-flow-guided refiner is introduced to refine the generated candidate profiles. The refiner uses Laplacian positional encoding to anchor latent node representations to the physical grid structure, and embeds AC power-flow equations together with correlation-alignment constraints into the refinement objective. Through residual correction, it improves numerical accuracy while preserving temporal fluctuations and spatial correlations.

The rest of this paper is organized as follows: In Section II, we introduce the methodology. We demonstrate the case study results in Section III and conclude the paper in Section IV.

## II. Methodology

### A. Problem Formulation

Consider a distribution network modeled as an undirected spatial graph $\mathcal{G} = (\mathcal{V}, \mathcal{E})$ comprising $N$ nodes. Define $\mathbf{X} \in \mathbb{R}^{T\times N\times 3}$ as a data sample over $T$ time steps where the three channels correspond to voltage $V$, active power $P$, and reactive power $Q$. For each time step $t$ and bus $n$

$$\mathbf{X}_{t,n,:} = \left[ V_{t,n}, P_{t,n}, Q_{t,n} \right]. \tag{1}$$

During practical grid operations, the data tensor frequently experiences random and unknown corruption. This results in missing data segments across both spatial and temporal dimensions. Let $\mathbf{X}_{\text{obs}}$ denote the available non-missing data in $\mathbf{X}$. Let $\mathbf{X}_{\text{m}}$ denote the collection of all missing data segments to be inpainted. The objective of this profile inpainting task is to find an optimal mapping function $f_\theta$ parameterized by $\theta$, which aims to estimate $\mathbf{X}_{\text{m}}$ based on $\mathbf{X}_{\text{obs}}$ and $\mathcal{G}$. So, the problem can be formulated as:

$$\tilde{\mathbf{X}}_{\text{m}} = f_\theta(\mathbf{X}_{\text{obs}}, \mathcal{G}) \tag{2}$$

where $\tilde{\mathbf{X}}_{\text{m}}$ is the inpainted data.

### B. Framework Overview

The proposed Flow-PIN framework performs profile inpainting on incomplete distribution network datasets through a continuous process consisting of three parts: data preprocessing, a conditional flow matching model, and a topology-aware power-flow-guided refiner.

Firstly, the raw active power, reactive power, and voltage measurements are organized into a spatial and temporal matrix. A binary mask matrix is concurrently defined to identify the missing data. Static graph features, such as topological and correlation graphs, are extracted to represent network characteristics. Secondly, following data preprocessing, the CFM model is deployed to learn the data distribution and generate high-quality candidate values. This generative process is conditioned on the static graphs and dynamic mask information. A dynamic weighting strategy is applied to the physical penalty within the loss function. This strategy progressively strengthens physical feasibility without disrupting the generative prior. Third, the TPR driven by a differentiable correlation alignment mechanism is introduced. The output of generative models can align with real statistical characteristics. While TPR fine-tunes the output to further improve accuracy, direct fine-tuning often disrupts established spatial correlations. Therefore, the proposed refiner leverages global self-attention mechanisms and relaxed AC power flow constraints to enhance point-wise numerical precision while maintaining global statistical correlations.

## C. Data Preprocessing

As shown in Fig. 1, data preprocessing establishes the numerical basis for the profile inpainting task. Real-world electrical dataset exhibit different numerical scales. Min-max normalization is applied to scale variables into a uniform range. The minimum parameter and the maximum are calculated strictly from the training dataset to prevent information leakage into the testing phase.

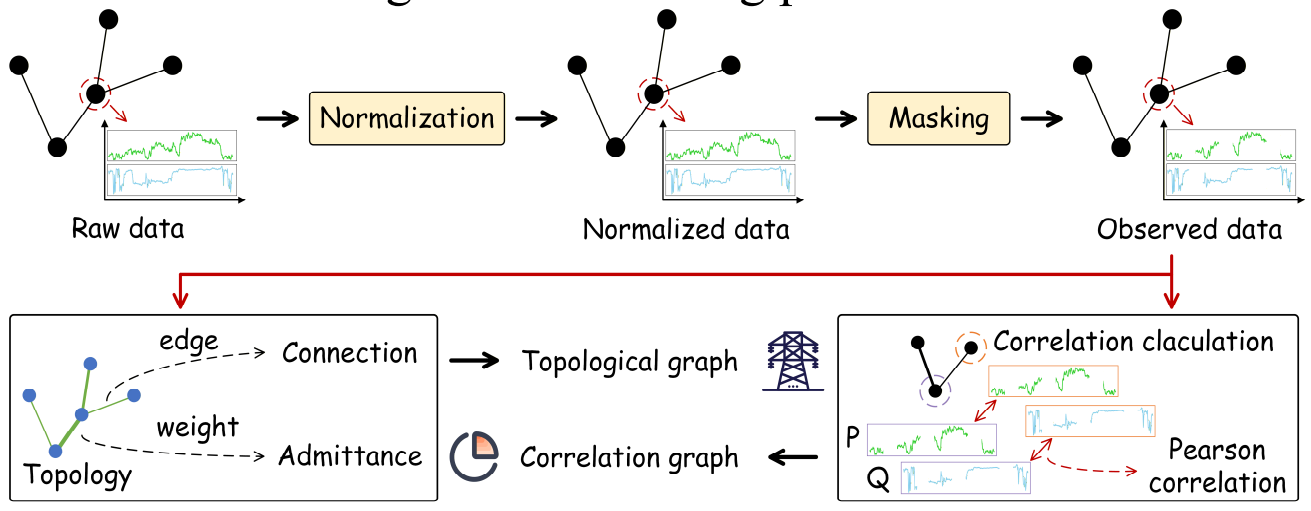

Fig. 1. The workflow of data preprocessing.

To simulate realistic missing values, a mixed spatio-temporal mask matrix $\mathbf{M} \in \{0,1\}^{T\times N\times 3}$ is constructed. A matrix element value of 1 indicates an available value and 0 indicates a missing one. The generation process simulates continuous time block failures rather than dropping random isolated points. The observed incomplete input data is defined as:

$$\mathbf{X}_{\text{obs}} = \mathbf{M} \odot \text{Norm}(\mathbf{X}), \tag{3}$$

where $\odot$ denotes the Hadamard product.

The spatial dependencies between nodes are modeled using a dual graph structure. The first component is the topological admittance graph $\mathbf{A}_{\text{topo}} \in \mathbb{R}^{N\times N}$. The edge weight $a_{\text{topo}}^{i,j}$ between node $i$ and node $j$ is calculated based on the branch admittance. Given the line resistance $R_{i,j}$ and reactance $X_{i,j}$, the element value is computed as:

$$a_{\text{topo}}^{i,j} = \frac{1}{\sqrt{R_{i,j}^2 + X_{i,j}^2}} \tag{4}$$

The resulting topological graph matrix is normalized by its maximum value and the diagonal elements are set to one.

The second component is the correlation graph $\mathbf{A}_{\text{cor}} \in \mathbb{R}^{N\times N}$. This graph captures the similarities between different nodes. The edge weights are formulated using the Pearson correlation coefficients of the active power and reactive power. Let $\rho_{\text{P}}^{i,j}$ and $\rho_{\text{Q}}^{i,j}$ denote the correlation coefficients between node $i$ and node $j$ for active power and reactive power respectively. The edge weight is formulated as:

$$a_{\text{cor}}^{i,j} = 0.5|\rho_{\text{P}}^{i,j}| + 0.5|\rho_{\text{Q}}^{i,j}| \tag{5}$$

A numerical threshold is applied to this matrix to remove weak correlations and the diagonal elements are set to one. This dual graph design ensures the model perceives both the hard topological connections and the soft statistical similarities within the network.

## D. Conditional Flow Matching Model

### 1) Conditional flow model formulation

A conditional flow matching model is employed to generate base inpainting estimates for the missing segments. Flow matching is a generative modeling method based on continuous normalizing flows. It constructs a probability path to transform a noisy distribution into the empirical distribution of the electrical variables. Let $\mathbf{x}_0$ represent a sample drawn from a Gaussian noise distribution. Let $\mathbf{x}_1$ represent the target ground truth. For a time step $t \in [0, 1]$, the conditional probability path $p(\mathbf{x}_t|\mathbf{x}_1)$ is defined as:

$$p(\mathbf{x}_t \mid \mathbf{x}_1) = \mathcal{N}(\mathbf{x}_t; t\mathbf{x}_1, (1-t)^2\mathbf{I}) \tag{6}$$

The $\mathbf{x}_t$ evolves along an optimal transport path, which is constructed through a linear interpolation formulation:

$$\mathbf{x}_t = t\mathbf{x}_1 + (1-t)\mathbf{x}_0 \tag{7}$$

Thus, the ground truth velocity field $\mathbf{v}_{\text{gt}}$ can be derived as:

$$\mathbf{v}_{\text{gt}} = \frac{d\mathbf{x}_t}{dt} = \mathbf{x}_1 - \mathbf{x}_0 \tag{8}$$

The generative process is governed by an Ordinary Differential Equation (ODE) defined by velocity estimator $v_\theta$, which is designed to approximate the ground truth velocity. The profile inpainting task is formulated as a conditional generation problem, where the network leverages contextual information surrounding the missing segments to achieve accurate profile inpainting. Specifically, by conditioning on diverse inputs, including current state $\mathbf{x}_t$, time $t$, and conditions, the network predicts the velocity field $\mathbf{v}_{\text{pred}}$.

$$\mathbf{v}_{\text{pred}} = v_\theta(\mathbf{x}_t, t| \; \mathbf{X}_{\text{obs}}, \mathbf{M}, \mathbf{A}_{\text{topo}}, \mathbf{A}_{\text{cor}}) \tag{9}$$

After the neural network is well-trained, the generative inference process is executed to generate missing segment estimates. Commencing from an initial noise distribution $\mathbf{x}_0 \sim \mathcal{N}(\mathbf{0}, \mathbf{I})$, the framework integrates the vector field $v_\theta$ across the continuous time interval from zero to one. The complete inpainting inference process is mathematically formalized as a definite integral by an ODE solver:

$$\mathbf{X}_{\text{gen}} = \mathbf{x}_0 + \int_0^1 v_\theta(\mathbf{x}_t, t| \; \mathbf{X}_{\text{obs}}, \mathbf{M}, \mathbf{A}_{\text{topo}}, \mathbf{A}_{\text{cor}}) \text{d}t \tag{10}$$

where the $\mathbf{X}_{\text{gen}}$ is the generative output. To guarantee precision at observable locations, the generated data are replaced by actual measurements utilizing the mask $\mathbf{M}$.

$$\mathbf{X}_{\text{base}} = \mathbf{M} \odot \mathbf{X}_{\text{obs}} + (1-\mathbf{M}) \odot \mathbf{X}_{\text{gen}} \tag{11}$$

This fused data $\mathbf{X}_{\text{base}}$ preserves known field observations while filling blind zones using the generative model.

### 2) Conditional velocity estimator

As shown in Fig. 2, the conditional vector field estimator is parameterized by a convolutional encoder-decoder U-Net architecture. To accurately model the spatiotemporal multivariate profiles, the estimator integrates both dynamic and static conditions.

First, the dynamic conditions comprise the observed incomplete data $\mathbf{X}_{\text{obs}}$ and the mask matrix $\mathbf{M}$. These dynamic matrices are directly concatenated with the noisy input data along the feature channel dimension. Concurrently, the continuous integration time is encoded to $\mathbf{t}_{\text{emb}}$ via a sinusoidal time embedding, processed through Multi-Layer Perceptrons (MLP), and injected into multiple residual blocks of the U-Net. This temporal conditioning enables the network to predict the appropriate vector field at each specific stage of the transformation.

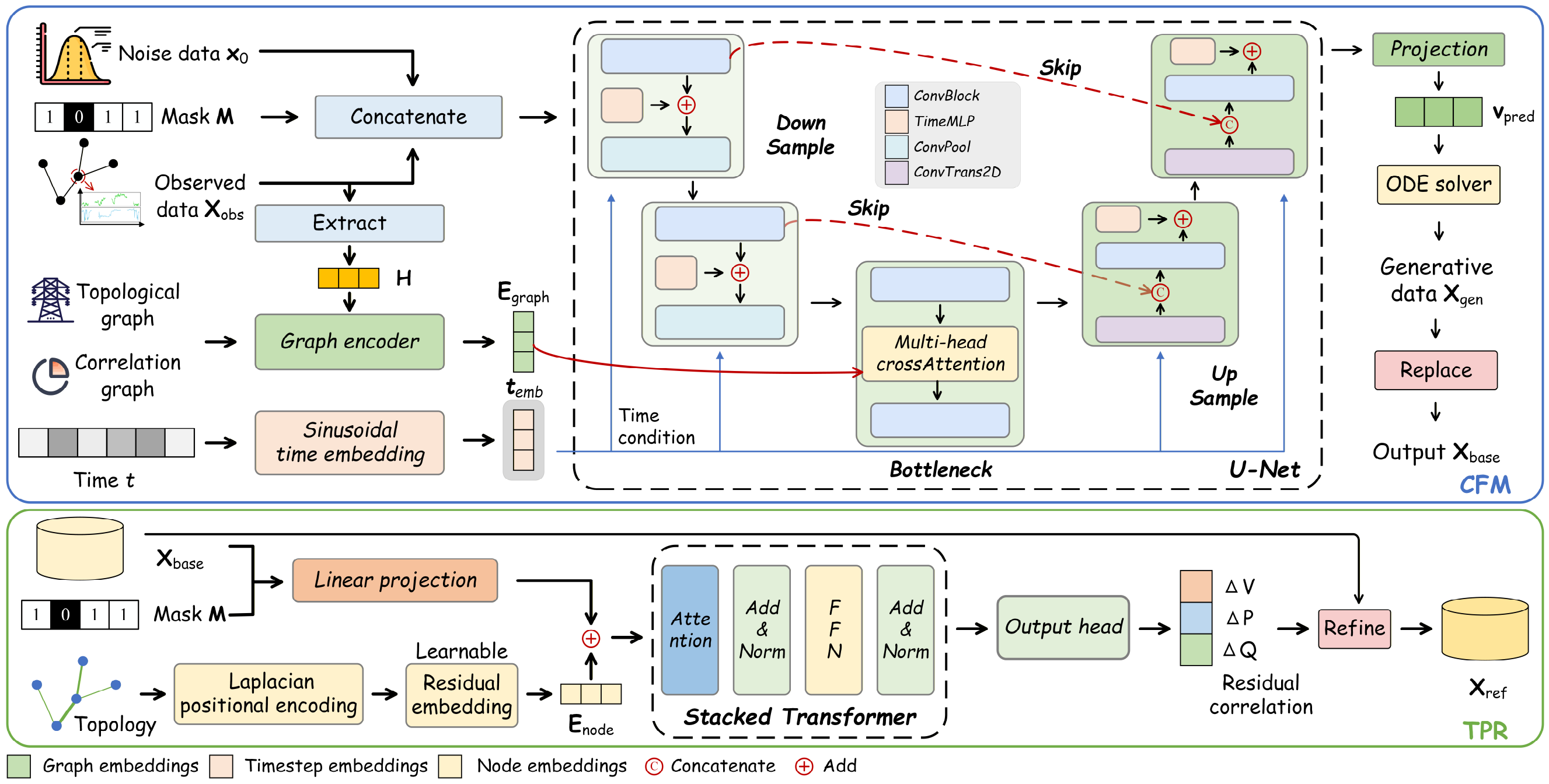

Fig. 2. The framework consists of two stages: the CFM module (top) and the TPR module (bottom). In the CFM stage, Gaussian noise, observed data, and the missingness mask are concatenated as inputs. Historical node features **H** are extracted from observations and processed by a graph encoder alongside dual-track priors (topological and correlation graphs) to form graph embeddings $\mathbf{E}_{\text{graph}}$. The $\mathbf{E}_{\text{graph}}$ is injected into the U-Net bottleneck via cross-attention, while sinusoidal time embeddings $\mathbf{t}_{\text{emb}}$ condition the network layers to predict the vector field $\mathbf{v}_{\text{pred}}$. An ODE solver is then utilized to generate continuous paths, followed by a hard replacement step at the observed locations to output the smoothed base inpainting estimate $\mathbf{X}_{\text{base}}$. In the TPR stage, $\mathbf{X}_{\text{base}}$ and the mask **M** are linearly projected and fused with topology-aware node embedding $\mathbf{E}_{\text{node}}$, which is constructed by superimposing learnable residuals onto Laplacian positional encodings. The fused representations are processed by a stacked Transformer encoder, and an output head generates residual corrections to align and refine the base data, yielding the final high-fidelity inpainted data $\mathbf{X}_{\text{ref}}$.

Second, the static conditions encompass the topological graph $\mathbf{A}_{\text{topo}}$ and the correlation graph $\mathbf{A}_{\text{cor}}$. To integrate these static structures, historical node features **H** are extracted by averaging the observed data $\mathbf{X}_{\text{obs}}$ along the temporal dimension. A unified graph embedding $\mathbf{E}_{\text{graph}}$ is computed through parallel linear transformations parameterized by weight matrices $\mathbf{W}_{\text{topo}}$ and $\mathbf{W}_{\text{cor}}$.

$$\mathbf{E}_{\text{graph}} = \text{ReLU}(\mathbf{A}_{\text{topo}}\mathbf{H}\mathbf{W}_{\text{topo}}) + \text{ReLU}(\mathbf{A}_{\text{cor}}\mathbf{H}\mathbf{W}_{\text{cor}}) \quad (12)$$

This graph embedding $\mathbf{E}_{\text{graph}}$ is introduced at the bottleneck of the U-Net architecture using a multi-head cross-attention mechanism. Specifically, the flattened spatiotemporal intermediate features **Z** generated by the convolutional encoder serve as the query, while $\mathbf{E}_{\text{graph}}$ serves as the key and value. The cross attention dynamically updates these latent features into a structure-aware representation $\mathbf{Z}_{\text{attn}}$ utilizing query weight $\mathbf{W}_{\text{Q}}$, key weight $\mathbf{W}_{\text{K}}$, and value weight $\mathbf{W}_{\text{V}}$ along with a scaling factor $d$. This process integrates the structural information of the distribution network into the vector estimator.

$$\mathbf{Z}_{\text{attn}} = \mathbf{Z} + \text{Softmax}\left(\frac{(\mathbf{Z}\mathbf{W}_{\text{Q}})(\mathbf{E}_{\text{graph}}\mathbf{W}_{\text{K}})^{\top}}{\sqrt{d}}\right)(\mathbf{E}_{\text{graph}}\mathbf{W}_{\text{V}}) \quad (13)$$

*3) Physics-guided composite loss function*

The neural network is optimized using a composite loss function to improve fidelity. The primary objective of the flow matching model is to regress the neural network output toward a target vector field. Thus, the base flow matching loss $\mathcal{L}_{\text{FM}}$ is formulated as the Mean Squared Error (MSE) between the predicted vector field $\mathbf{v}_{\text{pred}}$ and the target vector field $\mathbf{v}_{\text{gt}}$ over all time steps and data samples.

$$\mathcal{L}_{\text{FM}} = \mathbb{E}_{t,\mathbf{x}_0,\mathbf{x}_1}\left[\left\| v_\theta(\mathbf{x}_t, t \mid \mathbf{X}_{\text{obs}}, \mathbf{M}, \mathbf{A}_{\text{topo}}, \mathbf{A}_{\text{cor}}) - \mathbf{v}_{\text{gt}} \right\|_F^2\right] \quad (14)$$

During the training phase, the complete ground truth data **X** is accessible. A masked MSE loss $\mathcal{L}_{\text{MSE}}$, which focuses on the missing segments, is designed to enhance inpainted data precision. This loss is computed as:

$$\mathcal{L}_{\text{MSE}} = \mathbb{E}_{t,\mathbf{x}_0,\mathbf{x}_1}\left[\frac{\left\|(1-\mathbf{M}) \odot \left(\mathbf{v}_{\text{pred}} - \mathbf{v}_{\text{gt}}\right)\right\|_F^2}{\|1-\mathbf{M}\|_1 + \epsilon}\right] \quad (15)$$

where $\|1-\mathbf{M}\|_1$ represents the $L_1$ norm (i.e., the total count of missing elements) and $\varepsilon$ is a small positive constant.

Then, a physics-guided voltage drop penalty $\mathcal{L}_{\text{phys}}$ is integrated. Active distribution networks can contain radial feeders or meshed grids. A generalized topology agnostic linear power flow is formulated using a nodal impedance matrix. Let $\mathbf{R}_{\text{bus}}$ and $\mathbf{X}_{\text{bus}}$ represent the real and imaginary parts of the nodal impedance matrix excluding the slack bus. Let **P** and **Q** denote the generated active and reactive power matrices. Let $\mathbf{V}_{\text{pq}}$ denote the generated voltage at the corresponding nodes and $\mathbf{V}_{\text{slack}}$ denote the slack bus voltage. The theoretical voltage expectation is calculated through matrix multiplication. The physical loss $\mathcal{L}_{\text{phys}}$ is defined as:

$$\mathcal{L}_{\text{phys}} = \left\|\mathbf{V}_{\text{pq}} - (\mathbf{V}_{\text{slack}} - \mathbf{R}_{\text{bus}}\mathbf{P} - \mathbf{X}_{\text{bus}}\mathbf{Q})\right\|_F^2 \quad (16)$$

This differentiable formulation eliminates the need for complex path search algorithms. However, optimizing generative and physical objectives simultaneously often leads to gradient conflicts. To ensure stable mathematical optimization, the physical loss $\mathcal{L}_{\text{phys}}$ is governed by a dual-level dynamic weighting strategy.

The first level utilizes a continuous temporal coefficient. The raw physical loss is weighted by the squared time step $t^2$, forming the temporal-weighted physical loss $\mathcal{L}_{\text{phys}} = t^2 \mathcal{L}_{\text{phys_raw}}$. This temporal weight effectively suppresses the physical penalty during the initial generation phase ($t \to 0$) when the data are predominantly random noise, and rapidly amplifies the physical constraints as the generation nears completion ($t \to 1$).

The second level implements an adaptive gradient balancing mechanism during backpropagation, aiming to align the physical gradients with the generative gradients in magnitude. Let $\mathbf{W}_{\text{shared}}$ denote the weight of a shared output projection layer. At any given training iteration $k$, the network computes the $L_2$ gradient norm of the base generative losses $g_{\text{base}}^{(k)}$ and the physical loss $g_{\text{phys}}^{(k)}$:

$$g_{\text{base}}^{(k)} = \left\| \nabla_{\mathbf{W}_{\text{shared}}} (\mathcal{L}_{\text{FM}} + \lambda_{\text{mask}} \mathcal{L}_{\text{MSE}}) \right\|_2 \tag{17}$$

$$g_{\text{phys}}^{(k)} = \left\| \nabla_{\mathbf{W}_{\text{shared}}} \mathcal{L}_{\text{phys}} \right\|_2 \tag{18}$$

An instantaneous target ratio is calculated by dividing the base norm by the physical norm. To mitigate numerical oscillations across diverse data batches, an exponential moving average algorithm with a decay factor $\alpha$ is applied to compute the smoothed dynamic weight $\lambda_{\text{phys}}^{(k)}$:

$$\lambda_{\text{phys}}^{(k)} = \alpha \lambda_{\text{phys}}^{(k-1)} + (1-\alpha) \frac{G_{\text{base}}^{(k)}}{G_{\text{phys}}^{(k)} + \epsilon} \tag{19}$$

After being clipped within a predefined safety interval to improve training stability, this smoothed dynamic weight is integrated into the overall learning objective. The final backpropagation is executed using the dynamically balanced total loss:

$$\mathcal{L}_{\text{total}} = \mathcal{L}_{\text{FM}} + \lambda_{\text{mask}} \mathcal{L}_{\text{MSE}} + \lambda_{\text{phys}}^{(k)} \mathcal{L}_{\text{phys}} \tag{20}$$

where $\lambda_{\text{mask}}$ is a hyperparameter scaling the masked loss.

### *E. Topology-aware Power-flow-guided Refiner*

The generative model provides a high-fidelity base inpainting result that accurately reflects the statistical characteristics. Building upon this robust foundation, a subsequent fine-tuning stage is introduced to further elevate precision, thereby maximizing the ultimate profile inpainting quality. However, adjusting individual node values in isolation to meet local constraints can distort the spatial-temporal correlations established in the CFM stage.

To address this and maintain the fidelity of these correlations, the final stage of the proposed framework introduces the TPR module. As shown in Fig. 2, rather than treating node refinement as isolated independent processes, TPR leverages a global self-attention architecture to explicitly capture the complex physical coupling between distant buses. Guided by a composite optimization objective, this refiner seamlessly balances absolute numerical precision, physical constraints, and statistical correlation alignment.

#### *1) Residual-based refinement strategy*

To correct the initial predictions without completely overriding the distribution generated in the CFM stage, the TPR module adopts a residual learning strategy.

The refiner is mathematically defined as a mapping function $f_{\text{TPR}}$ that integrates the generative data $\mathbf{X}_{\text{base}}$, the mask matrix $\mathbf{M}$, and the network topology $\mathcal{G}$:

$$\{\Delta\mathbf{X}, \boldsymbol{\delta}\} = f_{\text{TPR}}(\mathbf{X}_{\text{base}}, \mathbf{M}, \mathcal{G}) \tag{21}$$

where $\Delta\mathbf{X}$ denotes the residual corrections and $\boldsymbol{\delta}$ denotes voltage phase angles used in the power-flow loss. Specifically, let $V_{\text{base},i}$, $P_{\text{base},i}$, and $Q_{\text{base},i}$ denote the initial candidate values generated by CFM for each node $i$. The refiner processes these inputs to compute their corresponding residual corrections $\Delta V_i$, $\Delta P_i$, $\Delta Q_i$, and a voltage phase angle $\delta_i$. The refined data are formulated by adding these corrections to the initial values:

$$V_{\text{ref},i} = V_{\text{base},i} + \Delta V_i, P_{\text{ref},i} = P_{\text{base},i} + \Delta P_i, Q_{\text{ref},i} = Q_{\text{base},i} + \Delta Q_i \tag{22}$$

where $V_{\text{ref},i}$, $P_{\text{ref},i}$, and $Q_{\text{ref},i}$ are the final refined data.

#### *2) Topology-aware self-attention architecture*

The TPR module employs a global self-attention architecture to capture the physical coupling across the distribution grid. The data processing pipeline consists of three stages: feature fusion, spatial embedding, and global feature extraction. Firstly, the base data $\mathbf{X}_{\text{base}}$ generated by CFM and the binary mask matrix $\mathbf{M}$ are concatenated along the feature dimension. This concatenated input is then mapped into a high-dimensional latent space via a linear projection layer.

To provide the model with explicit topological awareness, a hybrid topology-aware node embedding is introduced, which combines Laplacian Positional Encoding (LPE) with a learnable residual embedding. First, the Laplacian matrix $\mathbf{L}$ is computed based on the physical admittance graph of the distribution network. The eigenvectors are obtained by solving $\mathbf{L}\mathbf{u}_r = \lambda_r \mathbf{u}_r$, $r = 1, \ldots, N$. The eigenvectors associated with the $k$ smallest non-trivial eigenvalues are stacked as $\mathbf{U}_k = [\mathbf{u}_2, \mathbf{u}_3, \ldots, \mathbf{u}_{k+1}]$ and used as low-frequency structural priors. These eigenvectors are then linearly projected into the hidden dimension to form the topological baseline $\mathbf{E}_{\text{L}}$. To preserve the computational flexibility required for dynamic power flow variations, a learnable residual embedding $\mathbf{E}_{\text{res}}$ is initialized with marginal variance. The final topology-aware node representation $\mathbf{E}_{\text{node}}$ is formulated as:

$$\mathbf{L} = \mathbf{I} - \mathbf{D}^{-1/2} \mathbf{A} \mathbf{D}^{-1/2} \tag{23}$$

$$\mathbf{E}_{\text{node}} = \mathbf{E}_{\text{L}} + \mathbf{E}_{\text{res}} = \mathbf{U}_k \mathbf{W}_{\text{p}} + \mathbf{E}_{\text{res}} \tag{24}$$

where $\mathbf{I}$ is the identity matrix, $\mathbf{A}$ is the adjacency matrix, $\mathbf{D}$ is the diagonal degree matrix, and $\mathbf{W}_{\text{p}}$ represents the learnable linear projection weight matrix. The $\mathbf{E}_{\text{node}}$ is subsequently broadcasted across the temporal dimension and added directly to the projected data features before entering the transformer encoder.

The fused representations serve as the input to a stacked transformer encoder architecture, which consists of two identical layers. Within each layer, the data sequentially passes through a Multi-Head Self-Attention (MHSA) sub-layer and a position-wise Feed-Forward Network (FFN). Both sub-layers are accompanied by residual connections and layer normalization to prevent gradient degradation. The MHSA mechanism computes dense attention matrices across the flattened spatio-temporal sequences, dynamically capturing the physical coupling between distant buses. Subsequently, a feed-forward output head utilizes these processed features to generate the residual corrections $\Delta V_i$, $\Delta P_i$, $\Delta Q_i$, and a voltage phase angle $\delta_i$ for each node.

*3) Power-flow-guided and correlation-aligned loss function*

The training process is guided by four distinct objective functions to ensure accuracy and stability. Let Ω denote the set of node indices corresponding to the unobservable missing data locations. First, a data fidelity loss $\mathcal{L}_{\text{data}}$ is calculated. This loss measures the numerical deviation between the refined output and the ground truth data within the missing regions. It is formulated using specific weights $\alpha_v$ and $\alpha_{pq}$ for the voltage and power variables:

$$\mathcal{L}_{data} = \alpha_v \sum_{i\in\Omega} (V_{\text{ref},i} - V_{\text{gt},i})^2 + \alpha_{pq} \sum_{i\in\Omega} [(P_{\text{ref},i} - P_{\text{gt},i})^2 + (Q_{\text{ref},i} - Q_{\text{gt},i})^2] \tag{25}$$

where $V_{\text{gt},i}$, $P_{\text{gt},i}$, and $Q_{\text{gt},i}$ are the ground truth.

Second, voltage magnitudes are tightly regulated and generally less volatile than active and reactive power. To avoid excessive voltage drift during refinement, a relaxed voltage anchoring penalty $\mathcal{L}_{\text{anchor}}$ is introduced. This penalty focuses on the individual voltage adjustment amounts.

$$\mathcal{L}_{\text{anchor}} = \frac{1}{2} \sum_{i\in\Omega} (\Delta V_i)^2 \tag{26}$$

Third, to promote consistency with Kirchhoff's laws, a physical mismatch loss $\mathcal{L}_{\text{pf}}$ is implemented, which penalizes deviations between the refined injections and the AC power-flow equations. To facilitate numerical refinement without encountering ill-conditioned convergence, the refiner utilizes a rectangular coordinate system. The real component $e_i$ and imaginary component $f_i$ of the nodal voltages are formulated using the refined voltage and the learned phase angle $\delta_i$:

$$e_i = V_{\text{ref},i} \cos\delta_i, \quad f_i = V_{\text{ref},i} \sin\delta_i \tag{27}$$

Define $G_{ij}$ and $B_{ij}$ as the conductance and susceptance between node $i$ and node $j$ derived from the network admittance. For any node $i$, the theoretical active power $P_{\text{calc},i}$, reactive power $Q_{\text{calc},i}$ are calculated via AC power flow equations:

$$P_{\text{calc},i} = e_i \sum_j (G_{ij} e_j - B_{ij} f_j) + f_i \sum_j (G_{ij} f_j + B_{ij} e_j) \tag{28}$$

$$Q_{\text{calc},i} = f_i \sum_j (G_{ij} e_j - B_{ij} f_j) - e_i \sum_j (G_{ij} f_j + B_{ij} e_j) \tag{29}$$

The physical mismatch loss $\mathcal{L}_{\text{pf}}$ calculates the squared error between the network-refined power variables and these physically computed power data. Furthermore, because power transfer depends on relative phase angle differences, absolute phase angles possess infinite degrees of freedom, which can cause numerical drift. Thus, a regularization factor $\lambda_{\text{reg}}$ is applied to restrict the absolute magnitude of the phase angles:

$$\mathcal{L}_{\text{pf}} = \sum_{i\in\Omega} [(P_{\text{ref},i} - P_{\text{calc},i})^2 + (Q_{\text{ref},i} - Q_{\text{calc},i})^2] + \lambda_{\text{reg}} \sum_{i=1}^{N} \delta_i^2 \tag{30}$$

Finally, to preserve the statistical characteristics, a differentiable correlation alignment loss $\mathcal{L}_{\text{corr}}$ is implemented. This mechanism aligns the Pearson correlation of the refined variables with the ground truth distribution. This alignment focuses on the active power and reactive power. In power systems, individual voltages are regulated within a narrow range. This results in very low variance. This limited variance often leads to numerical instability during correlation calculations. In contrast, power variables exhibit high volatility and clear spatial dependencies driven by load behaviors. Let $\rho(\cdot)$ denote the differentiable Pearson correlation operator for the variable sequences across the batch. To function effectively within a deep learning optimization framework, an objective function cannot utilize standard statistical correlation algorithms. The calculation must maintain a continuous and differentiable computational graph to support backpropagation. Therefore, the operator $\rho(\cdot)$ is constructed entirely utilizing native tensor operations. This construction involves sequential mean subtraction, standard deviation division, and matrix multiplication. A marginal numerical constant is added to the standard deviation denominator to prevent division-by-zero errors. By satisfying these strict differentiability properties, the correlation alignment constraint seamlessly provides valid gradient updates to the self-attention mechanism.

To compute the numerical penalty between the correlation matrices, a Huber function $\mathcal{H}_\beta$ is utilized rather than a standard mean squared error. The Huber function acts as a squared error for small deviations and an absolute error for large deviations. This mathematical property prevents gradient explosion from numerical outliers while ensuring smooth optimization convergence. For a predicted data $x$ and a target value $y$, the Huber function with a threshold parameter $\beta$ is mathematically defined as follows

$$\mathcal{H}_\beta(x,y) = \begin{cases} \frac{1}{2\beta}(x-y)^2 & \text{if } |x-y| < \beta \\ |x-y| - \frac{1}{2}\beta & \text{otherwise} \end{cases} \tag{31}$$

Let $S$ {$P$, $Q$} represent the specific set of active and reactive power variables. The overall correlation alignment loss is compactly formulated by applying the Huber function to the derived correlation matrices of these power variables

$$\mathcal{L}_{\text{corr}} = \sum_{S\in\{P,Q\}} \mathcal{H}_\beta(\rho(S_{\text{ref}}), \rho(S_{\text{gt}})) \tag{32}$$

The total loss function for the refiner is formulated as:

$$\mathcal{L}_{\text{total}} = \mathcal{L}_{\text{data}} + \lambda_{\text{v}} \mathcal{L}_{\text{anchor}} + \lambda_{\text{pf}} \mathcal{L}_{\text{pf}} + \lambda_{\text{c}} \mathcal{L}_{\text{corr}} \tag{33}$$

where $\lambda_{\text{v}}$, $\lambda_{\text{pf}}$ and $\lambda_{\text{c}}$ are the weights to balance these components.

## III. CASE STUDY

### A. Experimental Setup

*1) Dataset*

To rigorously evaluate the proposed methodology under realistic operational conditions, a high-fidelity spatio-temporal dataset is constructed based on the standard IEEE 33-bus radial distribution system. The foundation of this dataset is built upon high-resolution real-world smart meter measurements from Pecan Street [20], which are high-quality 365-day residential load profiles. To accurately simulate the diversified load characteristics and mitigate individual stochastic spikes at the medium-voltage transformer level, a Monte Carlo aggregation strategy is employed by randomly sampling and superimposing ten individual residential profiles for each of the 32 load buses. Then, these aggregated profiles are downsampled to a 15-minute resolution, normalized against their 99.5th percentile values to prevent distortion from extreme outliers, and subsequently scaled to match the established rated active power specifications of the IEEE 33-bus topology. Furthermore, a 0.1% Gaussian white noise is injected into the scaled profiles to comprehensively emulate the underlying thermal noise typical of measurement sensors.

To reflect the inherent generation volatility of modern active distribution networks, DERs are strategically integrated into the system configuration, with solar Photovoltaic (PV) units assigned to buses 18 and 33, and wind generation units allocated to buses 22 and 25 [21], as shown in Fig. 3. Operating under a realistic assumption of a 0.95 lagging power factor for baseline loads and a 0.98 power factor for the distributed generators, the complete physical state of the grid is resolved through a continuous time-series non-linear AC power flow simulation covering an entire year, yielding 35,040 temporal steps. The Newton-Raphson algorithm is utilized to ensure robust convergence across the high-dimensional network manifold.

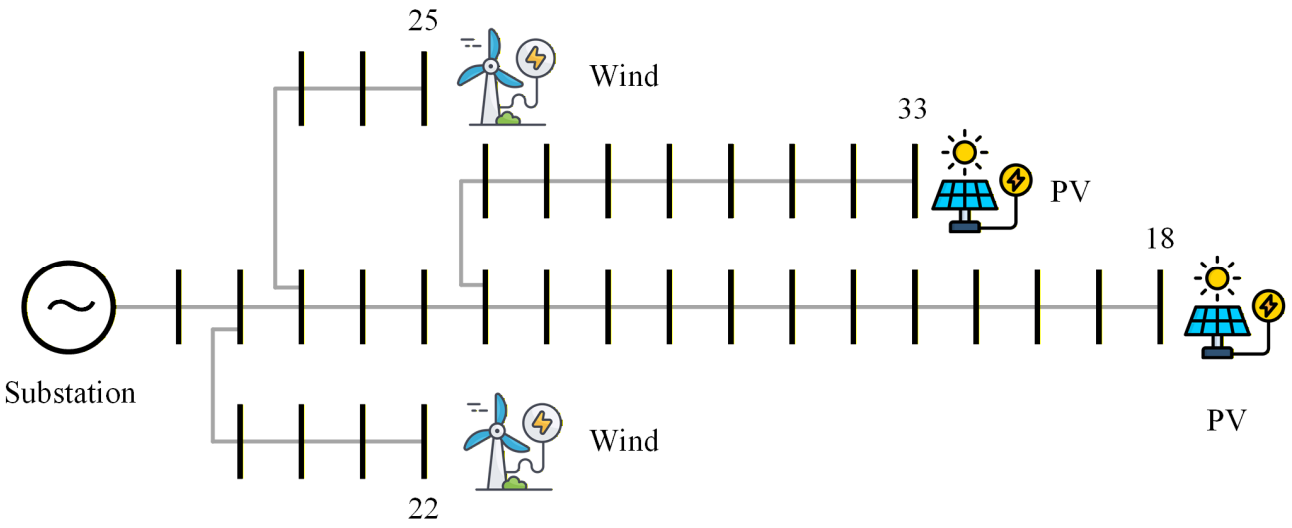

Fig. 3. The IEEE 33-bus distribution network.

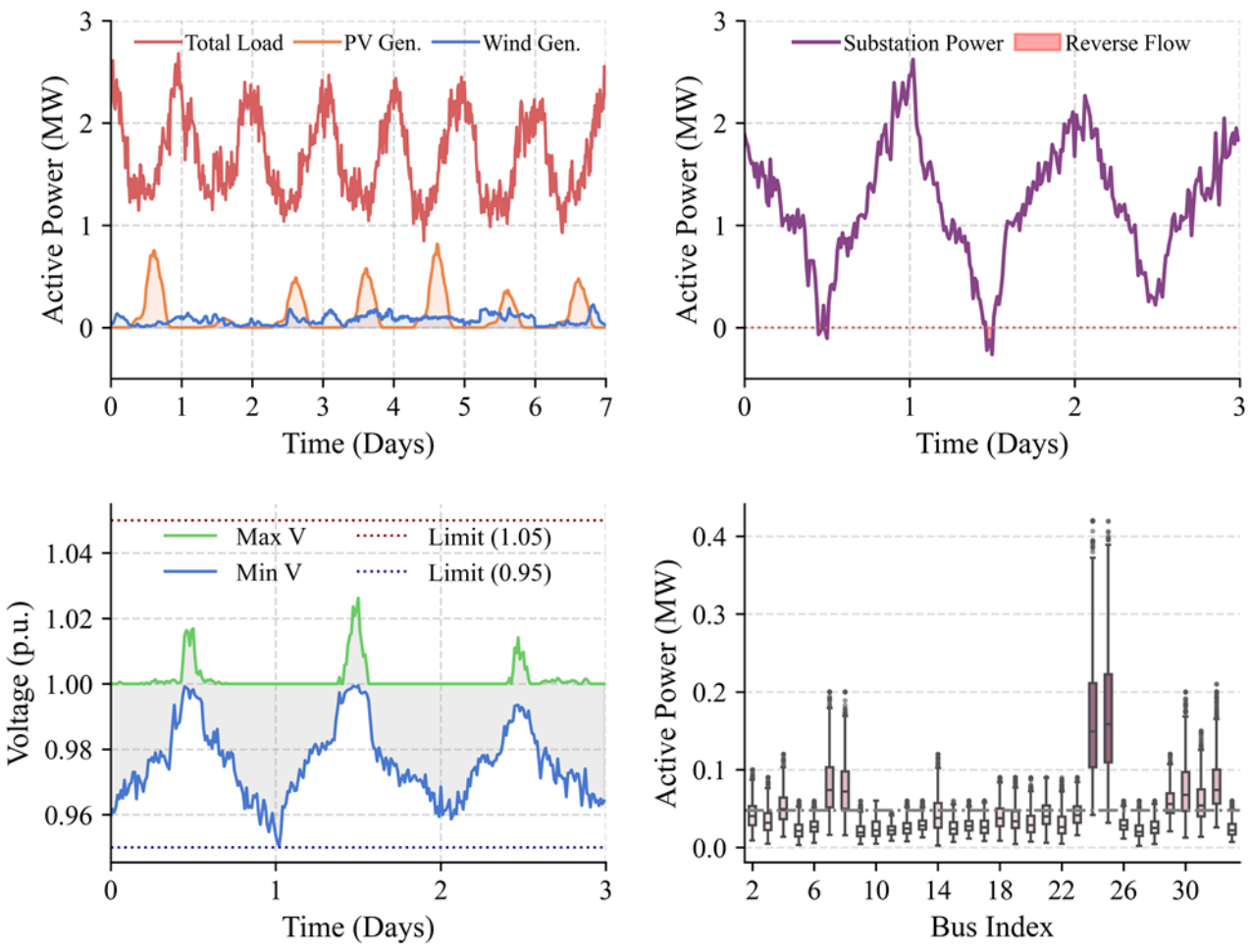

Fig. 4. Comprehensive spatiotemporal and physical validation of the synthesized IEEE 33-bus active distribution network dataset. (a) Representative weekly trajectories of the aggregated active power load alongside solar PV and wind generation. (b) Bidirectional active power exchange at the substation over a 3-day period, explicitly highlighting reverse power flow events during high-generation/low-load intervals. (c) Network-wide voltage envelope demonstrating the temporal voltage spread and boundary behaviors against the standard safety limits (0.95 p.u. and 1.05 p.u.). (d) Statistical distributions of the nodal active power demand across the 32 load buses, illustrating spatial heterogeneity and baseline capacity variations.

To empirically validate the representativeness and physical consistency of the synthesized dataset, a comprehensive spatiotemporal audit is conducted, as illustrated in Fig. 4. Fig. 4(a) depicts the representative weekly profiles of the aggregated load and DERs, explicitly capturing the complex net-load fluctuations. The presence of bidirectional power flow, a defining hallmark of active distribution network, is confirmed in Fig. 4(b). Specifically, the physical audit reveals a reverse power flow frequency of 0.30% across the year, demonstrating that high DER penetration during low-load intervals successfully forces power back to the substation grid. Furthermore, the spatiotemporal voltage envelope in Fig. 4(c) captures the characteristic voltage behavior of radial distribution networks. Under heavy-load conditions, the system exhibits end-of-line voltage drops and records 108 low-voltage events below 0.95 p.u. over the year. Simultaneously, the system maintains a tight global maximum phase angle spread of only 1.0507°. The limited maximum phase-angle spread indicates moderate angular variation under the simulated operating conditions. Together with the voltage-envelope and power-flow checks, this result supports the physical plausibility of the synthesized distribution-network states. Finally, Fig. 4(d) profiles the statistical capacity distribution for each load bus, verifying that the spatial heterogeneity successfully reflects the heterogeneous rated baseline configurations of the standard IEEE 33-bus topology while incorporating diverse real-world stochastic anomalies.

*2) Evaluation Metrics*

To evaluate the profile inpainting performance, we employ six metrics divided into three analytical perspectives. Let $K$ denote the total number of data points within the missing data segments $\mathbf{X}_{\mathrm{m}}$. The actual ground truth data at an index $k$ within $\mathbf{X}_{\mathrm{m}}$ is represented by $x_k^{\mathrm{m}}$, while $\tilde{x}_k^{\mathrm{m}}$ represents the corresponding inpainted value. To assess the point-wise accuracy and goodness-of-fit, we utilize Normalized Mean Absolute Error (NMAE), Normalized Root Mean Square Error (NRMSE), and Weighted Absolute Percentage Error (WAPE). They are formulated as:

$$\mathrm{NMAE} = \frac{1}{x_{\max} - x_{\min}} \frac{1}{K} \sum_{k=1}^{K} |\tilde{x}_k^{\mathrm{m}} - x_k^{\mathrm{m}}| \tag{34}$$

$$\mathrm{NRMSE} = \frac{1}{x_{\max} - x_{\min}} \sqrt{\frac{1}{K} \sum_{k=1}^{K} (\tilde{x}_k^{\mathrm{m}} - x_k^{\mathrm{m}})^2} \tag{35}$$

$$\mathrm{WAPE} = \sum_{k=1}^{K} |\tilde{x}_k^{\mathrm{m}} - x_k^{\mathrm{m}}| / \sum_{k=1}^{K} |x_k^{\mathrm{m}}| \tag{36}$$

These three metrics evaluate the numerical precision from different mathematical perspectives. NMAE measures normalized absolute errors linearly, providing a stable assessment of point-wise accuracy without disproportionate outlier penalties. In contrast, NRMSE penalizes large prediction deviations due to the squared term. The WAPE measures the total absolute deviation relative to the actual data volume. This avoids instability when ground truth values approach zero and reflects the overall reliability of the inpainted data.

Dynamic Time Warping (DTW) is introduced to evaluate the morphological alignment between the sequences. DTW searches for the optimal alignment path by minimizing the cumulative distance between $\mathbf{X}_{\mathrm{m}}$ and $\tilde{\mathbf{X}}_{\mathrm{m}}$. It is calculated as:

$$DTW(X_m, \tilde{X}_m) = D(K, K) \tag{37}$$

$$D(i,j) = |x_i^m - \tilde{x}_j^m| + \min \begin{cases} D(i-1, j) \\ D(i, j-1) \\ D(i-1, j-1) \end{cases} \tag{38}$$

where the distance matrix $D \in \mathbb{R}^{K\times K}$ is computed via dynamic programming. This metric evaluates the capability of the model in preserving the high-frequency temporal dynamics driven by renewable energy volatility.

Furthermore, it is crucial to assess the global distributional fidelity of the recovered data. For this purpose, we employ the Wasserstein Distance (WD) on the complete datasets. Let $\mathbb{P}$ and $\mathbb{Q}$ denote the underlying probability measures of the ground truth complete dataset and the recovered complete dataset, respectively. The WD is formally defined as:

$$W_1(\mathbb{P},\mathbb{Q}) = \inf_{\gamma\in\Pi(\mathbb{P},\mathbb{Q})} \mathbb{E}_{(x,y)\sim\gamma}[\|x-y\|] \quad (39)$$

where $\Pi(\mathbb{P}, \mathbb{Q})$ denotes the collection of all joint probability distributions whose marginals are $\mathbb{P}$ and $\mathbb{Q}$.

Beyond structural alignment, preserving high-frequency fluctuations is critical for renewable integration. Traditional regression methods often yield over-smoothed predictions by favoring mean trends. To mitigate this, we introduce the temporal volatility error, defined as the absolute difference between the mean absolute first-order differences of the recovered and ground truth sequences for a missing segment of length $L$:

$$\mathrm{TVE} = \left| \frac{1}{L-1}\sum_{l=1}^{L-1} |\tilde{x}_{l+1}^{\mathrm{m}} - \tilde{x}_{l}^{\mathrm{m}}| - \frac{1}{L-1}\sum_{l=1}^{L-1} |x_{l+1}^{\mathrm{m}} - x_{l}^{\mathrm{m}}| \right| \quad (40)$$

A value approaching zero signifies the successful reconstruction of realistic fluctuation dynamics.

Beyond point-wise accuracy and temporal dynamics, preserving the global spatial topological correlations among different nodes is critically essential for active distribution network analysis. To quantify this spatial structural fidelity, the Absolute Pearson Correlation Error (APCE) is evaluated. Let $\mathbf{C}_{\mathrm{gt}}$ and $\mathbf{C}_{\mathrm{rec}}$ denote the spatial Pearson correlation matrices using the ground truth and the recovered data, respectively. The spatial correlation error for any specific physical variable between a node pair $(i, j)$ is defined as the absolute mathematical deviation

$$E_{\mathrm{corr}}^{(i,j)} = \left| \left(\mathbf{C}_{\mathrm{rec}}\right)_{i,j} - \left(\mathbf{C}_{\mathrm{gt}}\right)_{i,j} \right| \quad (41)$$

A lower overall distribution of this absolute error confirms that the generative model successfully captures the intricate physical coupling and spatial interdependencies across the grid topology, successfully avoiding the over-smoothing degradation frequently observed in traditional regression baselines.

*3) Baselines*

To evaluate Flow-PIN, the complete framework is compared with nine external benchmark models and the base CFM variant. The baselines for temporal sequence modeling include the Bidirectional Long Short-Term Memory (Bi-LSTM) and the Gated Recurrent Unit (GRU). To incorporate recent structural developments in time series analysis, three general architectures are introduced: PatchTST based on patch-level attention [22], DLinear utilizing temporal trend decomposition [23], and TimesNet leveraging two-dimensional frequency domain convolutions [24]. Furthermore, the proposed method is compared with several deep generative and reconstruction architectures. These consist of a Variational Autoencoder (VAE), a Generative Adversarial Network framework (Load-PIN), a transformer-based masked language modeling approach (BERT-PIN), and a Conditional Score-based Diffusion Model (CSDI) [25]. Since Load-PIN and BERT-PIN are originally designed for univariate load profile restoration, their input channels are modified to properly accommodate the multivariate nature of the dataset.

### *B. Performance Comparison*

To validate the effectiveness of the proposed Flow-PIN, the imputation performance is evaluated against ten representative baselines across three electrical variables. The quantitative results are summarized in Table I, and the spatial fidelity analysis is illustrated in Fig. 5. The evaluation demonstrates the advantages of Flow-PIN across multiple analytical dimensions.

As detailed in Table I, Flow-PIN achieves the best overall reconstruction performance among the compared methods. It obtains the lowest NMAE and WAPE for voltage, active power, and reactive power, and the lowest NRMSE for voltage and active power, indicating strong point-wise accuracy across different electrical variables. The WAPE values show clear variable-dependent differences. Voltage magnitudes fluctuate within a narrow operating range around the nominal value, leading to percentage errors below 1.5%. In contrast, active and reactive power profiles are more volatile and may contain values close to zero, which naturally amplifies percentage-based errors. Despite these variable-specific error characteristics, Flow-PIN consistently reduces reconstruction errors compared with representative baselines such as CFM and CSDI, demonstrating the benefit of the proposed power-flow-guided generation and refinement framework.

Beyond numerical precision, the challenge in distribution network data imputation lies in preserving the high-frequency temporal dynamics driven by renewable energy integrations. Regression models, such as Bi-LSTM, GRU, and PatchTST, favor predicting the smoothed mean trend to minimize point-wise squared penalties. Consequently, these baselines suffer from the over-smoothing problem and fail to reconstruct the saw-tooth volatility of wind and solar outputs. In contrast, the proposed Flow-PIN achieves the lowest DTW and TVE. The minimized TVE validates that Flow-PIN avoids connecting missing endpoints with smoothed lines, and successfully synthesizes the physical fluctuation data across the time domain.

Furthermore, power system data reconstruction requires preserving statistical distributions and spatial topological correlations. Statistically, Flow-PIN achieves the lowest WD, indicating that the probability density functions of the imputed segments align with the ground truth. Spatially, as illustrated by the APCE distributions in Fig. 5, baseline models such as VAE and BERT-PIN exhibit larger correlation errors and wider variances, reflecting limitations in capturing the spatial coupling of the grid topology. In contrast, Flow-PIN maintains the correlation error near zero, recording the lowest median error and smallest variance among all benchmarks. These results demonstrate that the physics-informed refinement and differentiable correlation alignment mechanism preserve spatial interdependencies.

### *C. Ablation Study*

To evaluate the contribution of each module, an ablation

TABLE I
MODEL PERFORMANCES

| Variables | Voltage (V) | | | | | | Active Power (P) | | | | | | Reactive Power (Q) | | | | | |
|---|---|---|---|---|---|---|---|---|---|---|---|---|---|---|---|---|---|---|
| Metrics | NMAE | NRMSE | WAPE | TVE | DTW | WD | NMAE | NRMSE | WAPE | TVE | DTW | WD | NMAE | NRMSE | WAPE | TVE | DTW | WD |
| Bi-LSTM | 0.2168 | 0.2304 | 1.31% | 0.0034 | 0.0029 | 0.0075 | 0.0166 | 0.0386 | 50.96% | 0.0134 | 0.0067 | 0.0252 | 0.0206 | 0.0440 | 61.63% | 0.0043 | 0.0036 | 0.0068 |
| GRU | 0.2279 | 0.2816 | 1.47% | 0.0046 | 0.0032 | 0.0082 | 0.0163 | 0.0399 | 52.79% | 0.0092 | 0.0076 | 0.0235 | 0.0203 | 0.0449 | 66.77% | 0.0039 | 0.0034 | 0.0118 |
| PatchTST | 0.1257 | 0.1682 | 0.81% | 0.0013 | 0.0017 | 0.0032 | 0.0141 | 0.0362 | 45.72% | 0.0126 | 0.0067 | 0.0244 | 0.0138 | 0.0374 | 45.59% | 0.0034 | 0.0024 | 0.0066 |
| DLinear | 0.1039 | 0.1326 | 0.67% | 0.0014 | 0.0014 | 0.0034 | 0.0170 | 0.0426 | 55.23% | 0.0077 | 0.0083 | 0.0212 | 0.0155 | 0.0356 | 50.88% | 0.0022 | 0.0028 | 0.0089 |
| TimesNet | 0.0855 | 0.1206 | 0.55% | 0.0017 | 0.0017 | 0.0038 | 0.0233 | 0.0371 | 65.65% | 0.0094 | 0.0108 | 0.0399 | 0.0192 | 0.0393 | 63.25% | 0.0031 | 0.0032 | 0.0107 |
| VAE | 0.1409 | 0.1804 | 0.97% | 0.0023 | 0.0019 | 0.0056 | 0.0165 | 0.0389 | 48.30% | 0.0142 | 0.0075 | 0.0227 | 0.0148 | 0.0351 | 50.41% | 0.0042 | 0.0027 | 0.0070 |
| Load-PIN | 0.1263 | 0.1585 | 0.81% | 0.0013 | 0.0016 | 0.0035 | 0.0141 | 0.0367 | 45.69% | 0.0116 | 0.0059 | 0.0223 | 0.0137 | 0.0325 | 40.25% | 0.0031 | 0.0023 | 0.0060 |
| BERT-PIN | 0.1274 | 0.1618 | 0.76% | 0.0016 | 0.0018 | 0.0045 | 0.0154 | 0.0373 | 47.90% | 0.0127 | 0.0064 | 0.0228 | 0.0138 | 0.0337 | 45.14% | 0.0040 | 0.0022 | 0.0062 |
| CSDI | 0.1268 | 0.1846 | 0.83% | 0.0018 | 0.0014 | 0.0039 | 0.0142 | 0.0361 | 46.37% | 0.0117 | 0.0065 | 0.0217 | 0.0122 | 0.0326 | 43.93% | 0.0041 | 0.0026 | 0.0057 |
| CFM | 0.0785 | 0.1154 | 0.51% | 0.0012 | 0.0012 | 0.0031 | 0.0147 | 0.0366 | 45.04% | 0.0058 | 0.0057 | 0.0210 | 0.0119 | 0.0311 | 39.32% | 0.0019 | 0.0017 | 0.0056 |
| Flow-PIN | **0.0768** | **0.1083** | **0.50%** | **0.0012** | **0.0010** | **0.0015** | **0.0137** | **0.0289** | **38.19%** | **0.0058** | **0.0047** | **0.0174** | **0.0099** | **0.0304** | **32.53%** | **0.0017** | **0.0016** | **0.0045** |

*Bold term indicates the best performance, while underlining term represents second best.

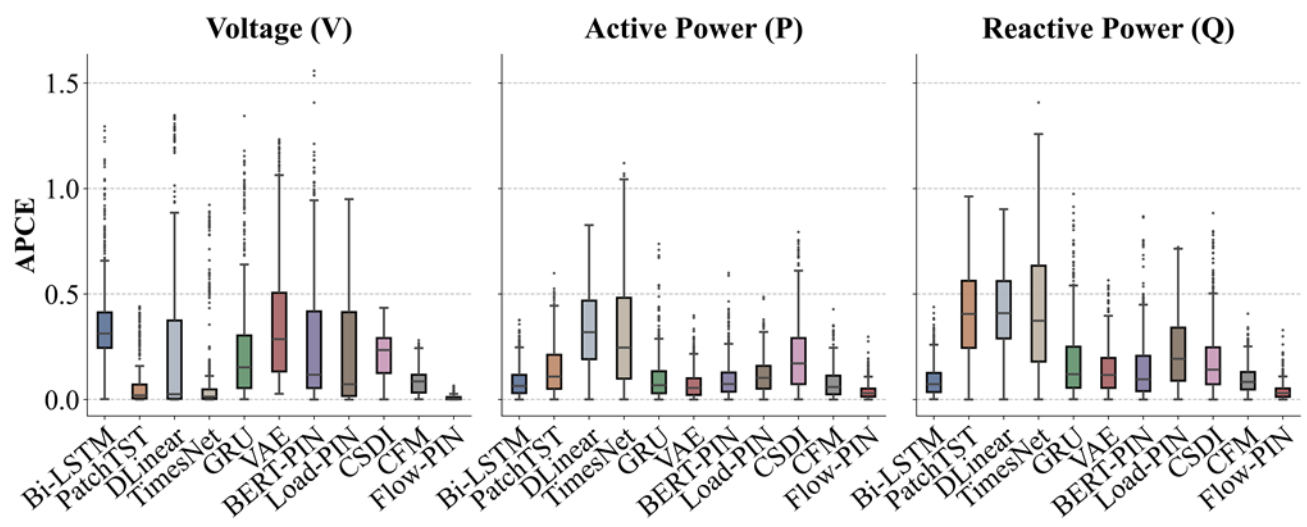

Fig. 5. Distribution of APCE for Voltage (V), Active Power (P), and Reactive Power (Q). The boxplots quantify the element-wise deviations between the reconstructed and ground-truth spatial correlation matrices across the network topology.

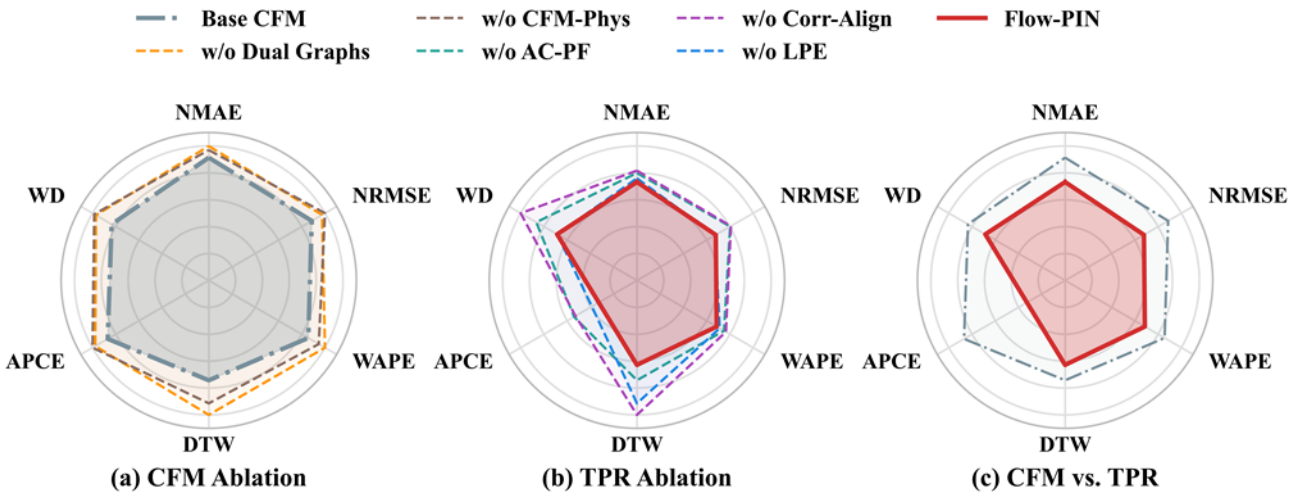

Fig. 6. Ablation results of Flow-PIN. The grouped radar charts compare the normalized errors of the complete model and its variants for the CFM stage, the TPR stage, and the complete two-stage framework.

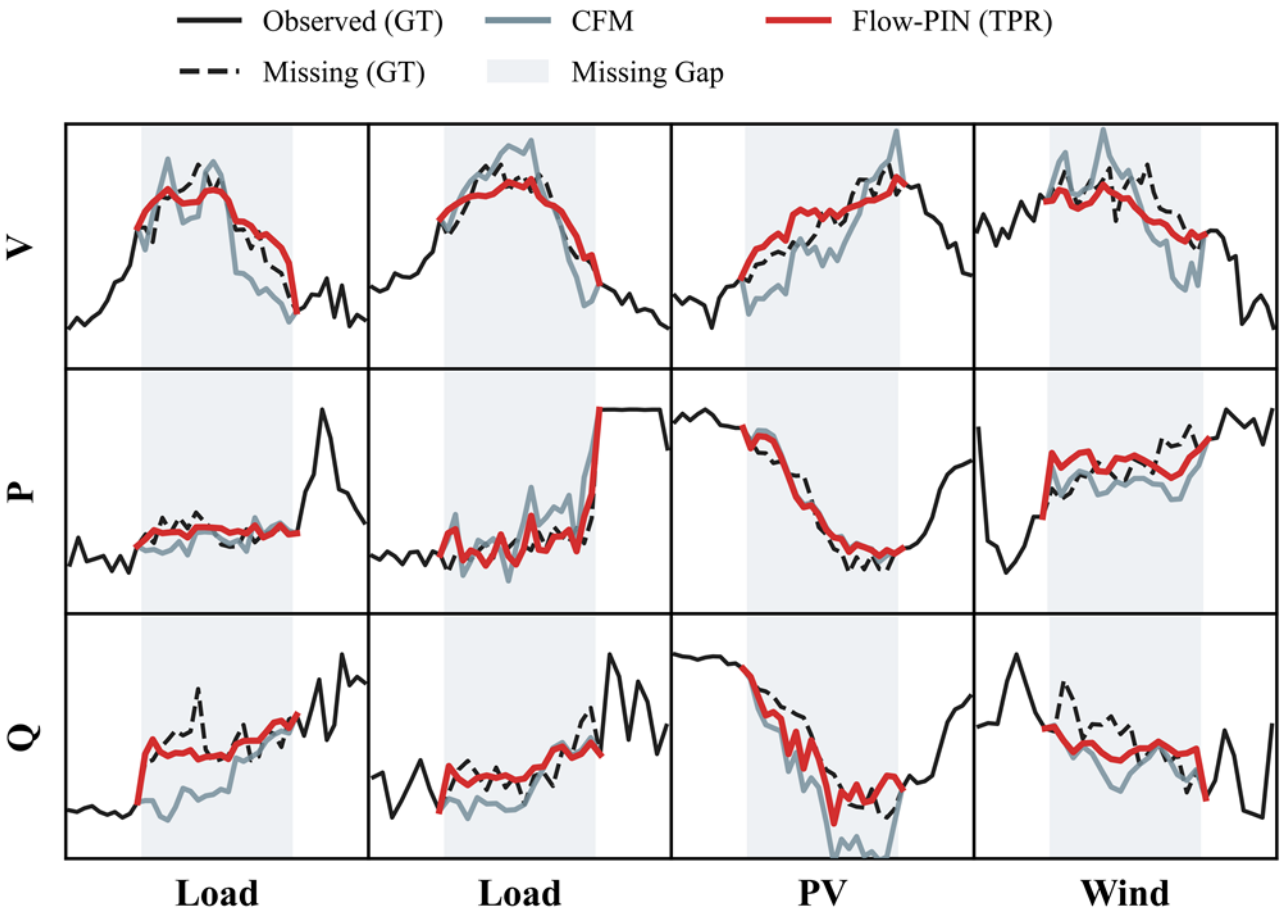

Fig. 7. Examples before and after applying TPR. This visualizes the inpainted curves for three variables across representative Load, PV, and Wind nodes. The grey shaded areas indicate the missing gaps. Compared to the base generative outputs (CFM, grey-blue lines), the refined data (Flow-PIN, red lines) exhibit reduced local deviations and closer alignment with the ground truth (black dashed lines).

study is conducted. Six model variants are constructed by removing components from the framework:

- **Base CFM:** The CFM model without the TPR.
- **w/o Dual Graphs:** Removes the topological graph and correlation graph conditions in CFM.
- **w/o CFM-Phys:** Disables the linear voltage drop penalty in the generative stage.
- **w/o LPE:** Removes the LPE within the TPR.
- **w/o AC-PF:** Removes the alternating current power flow mismatch loss during the refinement stage.
- **w/o Corr-Align:** Omits the correlation alignment mechanism during refinement.

The evaluation metrics of these variants are normalized and visualized in grouped radar charts in Fig. 6, with profile inpainting examples provided in Fig. 7.

Fig. 6(a) evaluates the components of CFM. Removing the dual graph conditions increases the APCE, indicating that topological and statistical conditions are necessary for capturing spatial coupling; without them, the generative model processes nodes in isolation. Furthermore, removing the physical penalty increases errors across all axes, demonstrating that linear physical priors steer the generative vector field toward a physically feasible domain.

Fig. 6(b) evaluates the TPR module. The w/o LPE variant expands the error envelope across numerical metrics such as NMAE and NRMSE. This highlights the necessity of explicitly injecting topological priors, as data-driven self-attention struggles to decode the underlying physical grid structure from measurement sequences. Additionally, removing the correlation alignment loss increases DTW and APCE, confirming that this mechanism preserves high-frequency dynamics and spatial dependencies. A trade-off is observed regarding the w/o AC-PF variant. While omitting the power-flow penalty yields lower point-wise numerical errors, the complete Flow-PIN framework minimizes APCE and WD. This result indicates that the AC-PF penalty does not merely pursue point-wise error reduction; instead, it guides the refiner toward power-flow-consistent and spatially coherent profiles, trading a small loss in local numerical accuracy for improved global distributional and topological fidelity.

To validate the necessity of the LPE, latent space alignments are compared between the complete TPR module and the w/o LPE variant. Randomly initialized node embeddings in the w/o LPE variant do not capture the grid structure, exhibiting near-zero geometric alignments (Spearman $\rho_{sp}$= -0.0173, Kendall $\tau$ = -0.0116, Pearson $r$= -0.0010, $p$ > 0.6). Without explicit structural priors, the attention mechanism clusters nodes based on transient operational similarities rather than physical connections, rendering the data-driven model topology-blind. In contrast, introducing LPE anchors the latent representations to the physical grid. The complete framework achieves statistically significant positive alignments, including a Spearman correlation of $\rho_{sp} = 0.1962$ ($p = 5.55 \times 10^{-6}$), a Kendall correlation of $\tau = 0.1347$ ($p = 4.11 \times 10^{-6}$) and a Pearson correlation of $r = 0.3497$ ($p = 1.25 \times 10^{-16}$). This comparison confirms that injecting graph priors guides the transformer toward physically meaningful representation learning.

Fig. 6(c) compares the Base CFM with the complete Flow-PIN framework. Flow-PIN exhibits a smaller error envelope across all metrics, validating that the TPR module bridges the gap between point-wise precision and statistical distribution. This result is corroborated by the inpainting examples in Fig. 7, which display curves for voltage, active power, and reactive power across load and generation nodes. Within the missing gaps, CFM approximates the trend but presents localized deviations. The refinement module corrects these deviations and reproduces the volatility of renewable energy generation. The two-stage framework integrates generative modeling with physical calibration to improve data fidelity.

## IV. Conclusion

This paper proposed a two-stage power-flow-guided generative framework for system-wide multivariate profile inpainting. By combining conditional flow matching with a topology-aware power-flow-guided refiner, Flow-PIN first learns graph-conditioned profile distributions and then calibrates the generated candidates using Laplacian positional encoding, AC power-flow constraints, and correlation alignment. Experiments against ten representative baselines show that Flow-PIN consistently improves profile inpainting across voltage, active power, and reactive power. The quantitative results indicate gains in both point-wise accuracy and distributional fidelity. It obtains the lowest NMAE for voltage, active power, and reactive power, with values of 0.0768, 0.0137, and 0.0099, respectively. Compared with the base CFM model, it further reduces WD by 51.6%, 17.1%, and 19.6%, indicating that the refinement stage improves distributional fidelity while preserving numerical accuracy.

Future work will focus on extending Flow-PIN to online profile inpainting under topology changes and abnormal operating conditions.